\documentclass[english]{extarticle}
\usepackage{courier}
\usepackage[T1]{fontenc}
\usepackage[latin9]{inputenc}
\usepackage{geometry}
\usepackage{array}
\usepackage{float}
\usepackage{units}
\usepackage{multirow}
\usepackage{amsmath}
\usepackage{graphicx}
\usepackage{wasysym}
\usepackage[authoryear]{natbib}

\makeatletter

\providecommand{\tabularnewline}{\\}

\newcommand{\lyxaddress}[1]{
	\par {\raggedright #1
	\vspace{1.4em}
	\noindent\par}
}

\usepackage{refstyle}
\usepackage[ruled]{algorithm2e}

\newref{fig}{%
    name      = \RSFigtxt,  
    names     = \RSFigstxt,
    Name      = \RSFigtxt,
    Names     = \RSFigstxt,
    rngtxt    = \RSrngtxt,
    lsttwotxt = \RSlsttwotxt,
    lsttxt    = \RSlsttxt}

\date{}

\makeatother

\usepackage{babel}
\begin{document}
\title{Efficient Particle Smoothing for Bayesian Inference in Dynamic Survival
Models}
\author{Parfait Munezero}
\maketitle

\lyxaddress{{\small{}Department of Statistics, Stockholm University}{\footnotesize{}}\\
{\footnotesize{}Parfait.Munezero@stat.su.se}}
\begin{abstract}
This article proposes an efficient Bayesian inference for piecewise
exponential hazard (PEH) models, which allow the effect of a covariate
on the survival time to vary over time. The proposed inference methodology
is based on a particle smoothing (PS) algorithm that depends on three
particle filters. Efficient proposal (importance) distributions for
the particle filters tailored to the nature of survival data and PEH
models are developed using the Laplace approximation of the posterior
distribution and linear Bayes theory. The algorithm is applied to
both simulated and real data, and the results show that it generates
an effective sample size that is more than two orders of magnitude
larger than a state-of-the-art MCMC sampler for the same computing
time, and scales well in high-dimensional and relatively large data.

\bigskip{}

\textbf{Key words}: Hazard function, Linear Bayes, particle filter,
particle smoothing, piecewise exponential, Survival function.
\end{abstract}

\section{Introduction}

The standard model for analysing survival data is the proportional
hazards model \Citep{david1972regression} which specifies the hazard
function as a product of a baseline hazard (an unknown function of
time, $t$) and a relative hazard (a function of the covariate vector,
$\mathbf{x}$). This model assumes that the ratio of the hazards corresponding
to two different covariate profiles is constant over time. However,
in some situations the effect of the covariate may change over time,
especially when the observation period is long. 

Piecewise exponential hazard (PEH) models \citep{gamerman1991dynamic}
allow the effect of the covariate to vary over time by assuming that
both the baseline and relative hazard functions are piecewise constant
over a set of disjoint and consecutive time intervals which partition
the observation period. PEH models are special cases of the piecewise
linear hazards models of \citet{murray2016flexible} and the more
general P-spline hazard models proposed by \citet{fahrmeir2011bayesian}.
Their key features are that the likelihood is tractable, and they
are flexible enough to capture various shapes of the hazard function.
Furthermore, they can be applied to both continuous survival time
\citep{wagner2011bayesian} and discrete survival time \citep{fahrmeir1996smoothing}.

Current Bayesian inference procedures for PEH models make use of Markov
Chain Monte Carlo (MCMC) methods. \citet{hemming2002parametric} build
on the transformation of the prior process suggested by \citet{gamerman1998markov}
to design a Metropolis-Hastings (MH) algorithm with random walk proposals.
\citet{fahrmeir2011bayesian} develop a MH algorithm with a multivariate
Gaussian proposal defined using the score function and observed information
matrix. \citet{wagner2011bayesian} develops a Gibbs sampler based
on the data augmentation method of \citet{fruhwirth1994data} and
proposes a variable selection scheme. The major drawback of these
methods is that they are computationally expensive as they require
many iterations to achieve convergence. Their convergence is hindered
by the auto-correlation of the effect parameters induced by the random
walk prior process. 

The aim of this paper is to propose a fast and efficient alternative
inference methodology for PEH models. The PS methodology is based
on Sequential Monte Carlo (SMC) methods, commonly known as particle
smoothing algorithms \citep*{briers2010smoothing,fearnhead2010sequential}.
These algorithms are specifically designed to sample effectively from
state-space models, of which PEH models is one instance. The main
advantages of PS algorithms are that: i) they require no functional
approximation of the likelihood, ii) they break a multidimensional
problem into sequences of smaller dimensional problems. This makes
them computationally fast and highly efficient especially when the
number of intervals partitioning the study period is relatively large. 

To be more specific, I apply the PS algorithm of \citet{fearnhead2010sequential},
which relies on three particle filtering approximations: a forward
particle filter which propagates information forwards in time, a backward
particle filter which propagates information backwards in time, and
a particle filter which combines the former two particle filters.
These algorithms require a proposal distribution (importance distribution)
that is easy to sample from. The so called bootstrap particle filters
propose particles (parameter draws) from the prior distribution. This
approach is prone to high degeneracy of the particles because the
proposal does not incorporate any evidence from data. The standard
way of incorportating evidence from data in the proposal distribution
relies on the second order Taylor series expension of the log likelihood
\citep{fearnhead2010sequential}. However, this is not appropriate
for the PEH models because the mode and the hessian of the likelihood
are not finite for censored individuals.

The main contribution of this paper is an efficient class of proposal
distributions specially adapted to the nature of survival data and
PEH models. The proposal of the forward filter is designed using the
Laplace approximation of the posterior distribution of the hazard
function with respect to the linear predictor (the log of the hazard
function) and the linear Bayes method of \citet{west1985dynamic}.
The proposal distributions of the other filters follow the linear
Bayes theory. Thus, the proposed inference algorithm is referred to
as the particle smoother with linear Bayes proposals (PSLiB). The
PSLiB algorithm is straightforward to implement since one needs only
the Laplace approximation of the posterior of hazard function, and
it can be easily extended to the class of multi-parameter regression
models for survival data \citep{burke2017multi} which model the survival
time parametrically with each distributional parameter allowed to
depend a set of covariates. Here, the only requirements would be the
Laplace approximation of the posterior distribution of the distributional
parameters with respect to the linear predictors. 

The proposed inference methodology is applied to both simulated and
real data. The simulation study, presented in Section \ref{sec:Applications},
aims at investigating the performance of PSLiB with respect to the
dimension of the covariate vector, the size of the dataset, the proportion
of censored observations in the data, and the length of the study
period. Results show that PSLiB is highly efficient in high-dimensional
data and fairly large data with many observations, and scales well
computationally. However, the performance degrades as the number of
covariates and the proportion of censored observations increase. Similar
performance degradation with respect to the number of covariates in
the model has been noted by \citet{villani2012generalized} in a different
class of models. Further, the comparison of PSLiB and the auxiliary
mixture sampler (AMS) of \citet{wagner2011bayesian} shows that PSLiB
outperforms AMS in terms of both the effective sample size and the
computation time.

The rest of the paper is organized as follows. The next section delineates
the likelihood and prior for the dynamic survival model. Section 3
presents the proposed inference methodology and, in Section 4, the
performance of PSLiB is assessed through simulated and observed data.
Finally, some concluding remarks and suggestions for future research
are provided in Section \ref{sec:Concluding-remarks}.

\section{Dynamic survival model specification\label{sec:Dynamic-Survival-Models}}

Let $\tilde{T}$ denote a random variable representing the survival
time, which is the time until an event of interest occurs. Usually
the survival time is not observed for all individuals participating
in the study: some individuals are lost before they have experienced
the event or, for some, the study ends before they experience the
event. Individuals whose survival time is not known are called (right)
\textit{censored} observations. Letting the random variable $C$ denote
the censoring time, the observed survival time is represented by the
random variable $T=\min(\tilde{T},C)$. 

The \textit{hazard function} describes the instantaneous rate at which
the event occurs, and it can be linked to the covariate vector $\mathbf{x}$
in various ways. This paper considers the Cox-type \Citep{david1972regression}
multiplicative hazard model 
\begin{equation}
\lambda(t|\mathbf{x})=\lambda_{0}(t)\exp\left(\mathbf{x}^{\prime}\mathbf{\boldsymbol{\beta}}\left(t\right)\right),\label{eq:dynamic model}
\end{equation}
where $\boldsymbol{\beta}\left(t\right)$ is a vector of time-varying
regression coefficients which models the effect of the covariates
on the hazard function. The \textit{survival function},

\begin{equation}
S\left(t|\mathbf{x}\right)=\exp\left(-\int_{0}^{t}\lambda(s|\mathbf{x})ds\right),\label{survival function}
\end{equation}
 is the probability that an individual with profile $\mathbf{x}$
has not experienced the event by time $t$. Given observed data for
$n$ individuals: the exposure time $t_{i}$, the censoring indicator
$d_{i}$ ($d_{i}=0$ if censored, and $d_{i}=1$ if event occurs),
and the covariate vector $\mathbf{x}_{i}$ (for $i=1,\cdots,n$),
the likelihood function is expressed as

\begin{equation}
L\left(t_{1},\ldots,t_{n}\vert\mathbf{\boldsymbol{\beta}}\left(t\right)\right)=\prod_{i=1}^{n}\lambda\left(t_{i}|\mathbf{x}_{i}\right)^{d_{i}}S\left(t_{i}|\mathbf{x}_{i}\right).\label{eq:likelihood}
\end{equation}

This model can be factorized sequentially into temporal factors by
assuming that the hazard function is piecewise constant, which results
in the PEH model \citep{gamerman1991dynamic}.

\subsection{The piecewise exponential hazard model\label{subsec:PE model}}

The PEH model partitions time into consecutive disjoint intervals,
$I_{j}=[\tau_{j-1},\tau_{j})$ (where $j=1,\cdots,J$ and $\tau_{0}=0<\tau_{1}<,\cdots,<\tau_{J}$),
and assumes that the baseline hazard function is constant within each
interval $I_{j}$ ; i.e. $\lambda_{0}(t)=\lambda_{0j}$, for $t\in I_{j}$
and $\lambda_{0j}>0$. Furthermore, it assumes that the vector of
regression coeficients $\boldsymbol{\beta}\left(t\right)$ is piecewise
constant; i.e $\boldsymbol{\beta}\left(t\right)=\boldsymbol{\beta}_{j}$
if $t\in I_{j}$. Therefore, the hazard function is represented by
several constant parameters $\lambda_{1},\ldots,\lambda_{J}$, where
each $\lambda_{j}$ is connected to the covariate information of an
individual $i$ through the log link

\begin{equation}
\ln\lambda_{ij}=\mathbf{z}_{i}^{\prime}\mathbf{\boldsymbol{\beta}}_{j},\label{eq:lambda}
\end{equation}
which allows the flexibility to capture different shapes of the hazard
function across time. Here, $\mathbf{z}_{i}=(1,\mathbf{x}_{i}^{\prime})$
is the original covariate vector of length $P$ augmented with a column
of $1$, $\mathbf{\boldsymbol{\beta}}_{j}=(\mathbf{\beta}_{0j},\mathbf{\beta}_{1j},\cdots,\mathbf{\beta}_{Pj})$
represents the vector of regression coefficients, where the intercept
$\mathbf{\beta}_{0j}=\ln(\lambda_{0j})$ is the log of the baseline
hazard. 

Partitioning time into discrete intervals generates interval-based
data. The survival time $t_{i}$ breaks into several exposure times
$t_{ij}=\max(0,\min(t_{i}-\tau_{j-1},\tau_{j}-\tau_{j-1}))$ which
define the amount of time an individual $i$ is exposed to the occurence
of the event during the interval $I_{j}$. The exposure time is equal
to the length of $I_{j}$ (if individual $i$ survived through this
interval), or it is equal to $t_{i}-\tau_{j-1}$ (if individual $i$
experienced the event or is censored in the interval $I_{j}$), otherwise
it is equal to zero. Furthermore, the event indicator expands into
a vector of binary variables $d_{ij}=1$ if the event occurs in interval
$I_{j}$, and $d_{ij}=0$ if the individual is censored in or survives
through the interval $I_{j}$.

With the assumption that covariates enter per (\ref{eq:lambda}),
the survival function for individual $i$ becomes

\begin{equation}
S\left(t_{i}|\boldsymbol{x}_{i}\right)=\exp\left(-\left[\sum_{j=1}^{h-1}\lambda_{ij}\left(\tau_{j}-\tau_{j-1}\right)\right]-\lambda_{ih}\left(t_{i}-\tau_{h-1}\right)\right),\,\textrm{if }\tau_{h-1}\leq t_{i}<\tau_{h},\,h\leq J,\label{eq:survival function}
\end{equation}
and the likelihood function (\ref{eq:likelihood}) can be factorized
across the intervals:
\begin{eqnarray}
L\left(\mathbf{t}_{1:J}\vert\boldsymbol{\beta}_{1:J}\right) & = & \prod_{j=1}^{J}L_{j}\left(\mathbf{t}_{j}|\boldsymbol{\beta}_{j}\right),\label{eq:interval likelihood}
\end{eqnarray}
where 
\begin{align*}
L_{j}\left(\mathbf{t}_{j}|\boldsymbol{\beta}_{j}\right) & =\prod_{i=1}^{n_{j}}\lambda_{ij}^{d_{ij}}\exp\left(-\lambda_{ij}t_{ij}\right),
\end{align*}
$\mathbf{t}_{j}$ is the vector of exposures for interval $I_{j}$,
$\mathbf{t}_{1:J}=(\mathbf{t}_{1},\ldots,\mathbf{t}_{J})$, $\boldsymbol{\beta}_{1:J}=(\beta_{1},\ldots,\beta_{J})$,
and $n_{j}$ is the number of individuals who experienced the event
during interval $I_{j}$. For more details and justification of the
likelihood given in (\ref{eq:interval likelihood}), see \citet{gamerman1991dynamic}.

Alternative likelihood expressions can be obtained via data augmentation
\citep{wagner2011bayesian} or partial likelihood \citep{sargent1997flexible}
approaches. The data augmentation approach completes the censored
exposure times (corresponding to $d_{ij}=0$) by a latent exponentially
distributed residual time. \citet{wagner2011bayesian} approximates
the log of the augmented survival times by a mixture of ten normal
components with known mean and variance parameters, resulting in a
dynamic linear model. This method is used in Section \ref{subsec:Simulations}
as a benchmark for assessing the performance of the inference methodology
proposed in this paper.

The PEH model relies on the method of partitioning the time into $J$
intervals. The partition can be based on the event times, placing
$\tau_{j}$ at each observed event time \citep{gamerman1991dynamic},
which may be computationally expensive especially when the number
of events is large. Alternatively, intervals could be equidistant
as in \citet{hemming2002parametric}, or they could contain equal
number of events per interval (in this way there are more intervals
in areas where there is more information). The selection of the number
of events per interval can be done using a model comparison measure
such as the Watanabe-Akaike information criterion (WAIC) descibed
in Section \ref{subsec:Model-comparison-and prediction}.

\subsection{Prior specification\label{subsec:Prior-specification}}

In order to complete the model specification, one needs to define
the prior process for the regression coefficients. One of the simplest
and most applied smoothing priors on $\mathbf{\boldsymbol{\beta}}_{j}$
is the random walk 

\begin{equation}
\mathbf{\boldsymbol{\beta}}_{j}=\mathbf{\boldsymbol{\beta}}_{j-1}+\epsilon_{j},\,\,\,\,\,\,\epsilon_{j}\sim N\left(0,\mathbf{U}_{j}\right).\label{eq:prior}
\end{equation}
This process is adopted by \citet{wagner2011bayesian}, \citet{fahrmeir1994dynamic}
and \citet{hemming2002parametric}, and is a special case of the more
general first order random walk process for parameter evolution suggested
by \citet{gamerman1991dynamic}. Clearly, if $\mathbf{U}_{j}$ is
a zero matrix then there is no change in the regression coefficients
and the dynamic model reduces to the standard proportional hazards
model. Otherwise, $\mathbf{\boldsymbol{\beta}}_{j}$ varies over time
and larger values for the entries in $\mathbf{U}_{j}$ indicate higher
variation in $\mathbf{\boldsymbol{\beta}}_{j}$.

In many applications,$\mathbf{U}_{j}$ is held constant (i.e, $\mathbf{U}_{j}=\mathbf{U}$)
and is assumed an unknown parameter. For a diagonal $\mathbf{U}_{j}$,
both lognormal \citep{hemming2002parametric} and inverse gamma \citep{sargent1997flexible,wagner2011bayesian}
priors have been suggested for the diagonal components. For a full
matrix $\mathbf{U}_{j}$, \citet{gamerman1998markov} assumes an inverse
Wishart prior on $\mathbf{U}_{j}$. Alternatively, \citet{west1985dynamic}
suggests a discounting procedure for approximating $\mathbf{U}_{j}$
in terms of a discount parameter $0<\phi<1$ that controls the amount
of information transferred through intervals. Given the posterior
variance $\boldsymbol{\Sigma}_{j-1}$ of the regression coefficients
in the interval $I_{j-1}$, the discount factor approach approximates
$\mathbf{U}_{j}=(\phi^{-1}-1)\boldsymbol{\Sigma}_{j-1}$. This allows
the variance $\mathbf{U}_{j}$ to vary over time, which improves the
capacity of the random walk process to adapt locally. It is also possible
to assume a time-varying discount factor (see \citet{das2013dynamic}).
However, in this paper, $\mathbf{U}_{j}$ is allowed to be a full
matrix, $\phi$ is held constant and inference on $\phi$ is based
on the WAIC described in Section \ref{subsec:Model-comparison-and prediction}.

\section{Inference}

This section describes the posterior distribution and the sequential
Monte Carlo sampling procedure used to sample from the target posterior
distribution,
\begin{equation}
p\left(\boldsymbol{\beta}_{1:J}|\mathbf{t}_{1:J}\right)\propto p\left(\mathbf{t}_{1}|\boldsymbol{\beta}_{1}\right)p\left(\boldsymbol{\beta}_{1}\right)\prod_{j=2}^{J}L_{j}\left(\mathbf{t}_{j}|\boldsymbol{\beta}_{j}\right)p\left(\boldsymbol{\beta}_{j}|\boldsymbol{\beta}_{j-1}\right),\label{eq:joint posterior}
\end{equation}
 where $p(\boldsymbol{\beta}_{j}|\boldsymbol{\beta}_{j-1})$ is defined
by the expression (\ref{eq:prior}). MCMC methods have been used in
the literature to sample from (\ref{eq:joint posterior}). \citet{hemming2002parametric}
reparametrize $\boldsymbol{\beta}_{j}$ in terms of the evolution
noise $\epsilon_{j}$ in (\ref{eq:prior}), and apply a Gibbs sampler
with a random walk Metropolis Hastings step for each $\boldsymbol{\beta}_{j}$.
\citet{wagner2011bayesian} designs a Gibbs sampler where the full
path $\boldsymbol{\beta}_{1:J}$ is sampled in one move using the
forward filtering backward sampling algorithm \citep{fruhwirth1994data}. 

Note that the likelihood in (\ref{eq:interval likelihood}) and the
random walk prior process (\ref{eq:prior}) define a state space model
with non-linear and non-Gaussian observation model \citep{gordon1993novel},
which allows to apply SMC inference methods. Generally, SMC methods
are specifically designed for filtering problems \citep{doucet2000sequential}
in state space models, where the main objective is to sample from
$p(\boldsymbol{\beta}_{1:j}\text{|}\mathbf{t}_{1:j}),\,j=1,\ldots,J,$
sequentialy through lower-dimensional marginals, $p(\boldsymbol{\beta}_{j}\text{|\,}\mathbf{t}_{1:j})$,
referred to as filtering distributions. 

SMC methods for sampling from (\ref{eq:joint posterior}) sequentially
through the smoothing distribution, $p(\boldsymbol{\beta}_{j}\text{|}\mathbf{t}_{1:J})$,
are also available in the literature. The forward-backward smoother
samples parameters from the filtering distribution and smooths them
in a backward procedure (see \citealp{doucet2000sequential} and references
therein). On the other hand, the two-filter smoother \citep{briers2010smoothing}
and its computationally cheaper variant \citep{fearnhead2010sequential}
combine samples from both a forward and a backward information filter.
This paper builds on the algorithm of the two-filter smoothing suggested
by \citet{fearnhead2010sequential} and the linear Bayes method \citep{west1985dynamic}
to design an efficient inference scheme for PEH models.

\subsection{The two-filter smoother\label{subsec:The-two-filter-smoother}}

The two-filter smothing recursion \citep{briers2010smoothing} reformulates
the smoothing distribution as
\begin{equation}
p\left(\boldsymbol{\beta}_{j}\text{|\,}\mathbf{t}_{1:J}\right)\propto p\left(\boldsymbol{\beta}_{j}\text{|\,}\mathbf{t}_{1:j-1}\right)p\left(\mathbf{t}_{j:J}|\,\boldsymbol{\beta}_{j}\right),\label{eq:Two-filter smoother}
\end{equation}
where the first term on the right hand side,
\begin{equation}
p\left(\boldsymbol{\beta}_{j}|\mathbf{t}_{1:j-1}\right)=\int p\left(\boldsymbol{\beta}_{j}|\boldsymbol{\beta}_{j-1}\right)p\left(\boldsymbol{\beta}_{j-1}|\mathbf{t}_{1:j-1}\right)d\boldsymbol{\beta}_{j-1},\label{eq:forward filter}
\end{equation}
 is the predictive prior recursion, and the second term, 
\begin{equation}
p\left(\mathbf{t}_{j:J}|\,\boldsymbol{\beta}_{j}\right)=L_{j}\left(\mathbf{t}_{j}|\boldsymbol{\beta}_{j}\right)\int p\left(\mathbf{t}_{(j+1):J}|\,\boldsymbol{\beta}_{j+1}\right)p\left(\boldsymbol{\beta}_{j+1}|\boldsymbol{\beta}_{j}\right)d\boldsymbol{\beta}_{j+1},\label{eq:Backward filter}
\end{equation}
is the likelihood recursion. Both the expressions (\ref{eq:forward filter})
and (\ref{eq:Backward filter}) are computed recursively through two
independent filters: i) the forward filter which estimates sequentially
the filtering distribution $p(\boldsymbol{\beta}_{j-1}|\mathbf{t}_{1:j-1})$
forward in time, and ii) the backward filter which evaluates the likelihood
recursion (\ref{eq:Backward filter}) by mirroring the forward filter
sequentially backward in time.

The forward filter begins from $j=1$ with a predifined initial distribution
$p(\boldsymbol{\beta}_{1})$, and then proceeds propagating information
forward in time through the usual posterior update: 
\begin{equation}
p\left(\boldsymbol{\beta}_{j}|\mathbf{t}_{1:j}\right)\propto L_{j}\left(\mathbf{t}_{j}|\boldsymbol{\beta}_{j}\right)p\left(\boldsymbol{\beta}_{j}|\mathbf{t}_{1:j-1}\right),\,j=2,\ldots,J.\label{eq:forward filter update}
\end{equation}

On the other hand, the formulation of the backward filter is not straightforward.
It turns out that (\ref{eq:Backward filter}) is not a density for
$\boldsymbol{\beta}_{j}$ and hence the integral may not be finite.
To mirror the forward filter and ensure that the integral in (\ref{eq:Backward filter})
is always finite, \citet{briers2010smoothing} introduce an artificial
prior $\gamma_{j}\left(\boldsymbol{\beta}_{j}\right)$, for $j=J,\ldots,1$,
so that the backward recursion expression becomes an artificial posterior;
i.e, 
\begin{align}
\tilde{p}\left(\beta_{j}|\mathbf{t}_{j:J}\right) & \propto p\left(\mathbf{t}_{j:J}|\,\boldsymbol{\beta}_{j}\right)\gamma_{j}\left(\boldsymbol{\beta}_{j}\right)\label{eq: backward filtering distribution}\\
 & \propto L_{j}\left(\mathbf{t}_{j}|\boldsymbol{\beta}_{j}\right)\gamma_{j}\left(\boldsymbol{\beta}_{j}\right)\int\frac{\tilde{p}\left(\boldsymbol{\beta}_{j+1}|\mathbf{t}_{(j+1):J}\right)}{\gamma_{j+1}\left(\boldsymbol{\beta}_{j+1}\right)}p\left(\boldsymbol{\beta}_{j+1}|\boldsymbol{\beta}_{j}\right)d\boldsymbol{\beta}_{j+1},\nonumber 
\end{align}
where the first term in the inner expression of the integral is simply
the likelihood $p(\mathbf{t}_{(j+1):J}|\,\boldsymbol{\beta}_{j+1})$.
The backward filter starts at $J$ with the prior distribution $\gamma_{J}\left(\boldsymbol{\beta}_{J}\right)$
and proceeds by evaluating (\ref{eq: backward filtering distribution})
recursively backward in time, for $j=J-1,\ldots,1$ .

The smoothing distribution $p(\boldsymbol{\beta}_{j}|\mathbf{t}_{1:J})$
is therefore estimated by combining the forward filter standing at
$j-1$ and the backward filter standing at $j+1$ (see Algorithm $1$).
If the likelihood is Gaussian and linear then both filters are analytically
tractable. Otherwise, some form of approximation such as the particle
filter approach \citep{doucet2000sequential,arulampalam2002tutorial}
is needed.

\subsubsection{Particle filtering approximation}

Particle filters provide a recursive procedure of approximating the
forward filtering distribution (\ref{eq:forward filter update}) by
an empirical distribution defined on a finite sample of points $\{\boldsymbol{\beta}_{j}^{k}\}_{k=1}^{K}$
(commonly known as particles) weighted by the probability masses $\{w_{j}^{k}\}_{k=1}^{K}$
(importance weights). Assuming that a sample of particles and their
corresponding importance weights at $j-1$ are available, then the
predictive prior (\ref{eq:forward filter}) is approximated as
\begin{equation}
p\left(\boldsymbol{\beta}_{j}|t_{1:j-1}\right)\propto\sum_{k=1}^{K}p\left(\boldsymbol{\beta}_{j}|\boldsymbol{\beta}_{j-1}^{k}\right)w_{j-1}^{k}.\label{eq:predictive prior approx}
\end{equation}

Similarly, the backward filtering distribution (\ref{eq: backward filtering distribution})
is approximated empirically by a finite sample of particles $\{\tilde{\boldsymbol{\beta}}_{j}^{k}\}_{k=1}^{K}$
weighted by $\{\tilde{w}_{j}^{k}\}_{k=1}^{K}$, and if the backward
filter stands at the time point $j+1$, then the likelihood (\ref{eq:Backward filter})
is approximated as
\begin{equation}
p\left(\mathbf{t}_{j:J}|\,\boldsymbol{\beta}_{j}\right)\propto L_{j}\left(\mathbf{t}_{j}|\boldsymbol{\beta}_{j}\right)\sum_{h=1}^{K}\frac{p\left(\tilde{\boldsymbol{\beta}}_{j+1}^{h}|\boldsymbol{\beta}_{j}\right)}{\gamma_{j+1}\left(\tilde{\boldsymbol{\beta}}_{j+1}^{h}\right)}\tilde{w}_{j+1}^{h}.\label{eq:Likelihood recurrsion approx}
\end{equation}

Therefore, realizations from the forward particle filter standing
at the time point $j-1$ and the backward filter at $j+1$ can be
combined to approximate the smoothing distribution (\ref{eq:Two-filter smoother})
as 
\begin{equation}
p\left(\boldsymbol{\beta}_{j}\text{|\,}\mathbf{t}_{1:J}\right)\propto\sum_{k=1}^{K}\sum_{h=1}^{K}p\left(\boldsymbol{\beta}_{j}|\boldsymbol{\beta}_{j-1}^{k}\right)L_{j}\left(\mathbf{t}_{j}|\boldsymbol{\beta}_{j}\right)\frac{p\left(\tilde{\boldsymbol{\beta}}_{j+1}^{h}|\boldsymbol{\beta}_{j}\right)}{\gamma_{j+1}\left(\tilde{\boldsymbol{\beta}}_{j+1}^{h}\right)}w_{j-1}^{k}\tilde{w}_{j+1}^{h}.\label{eq:  particle smoothing distribution}
\end{equation}

The posterior (\ref{eq:  particle smoothing distribution}) requires
estimates of the forward and backward importance weights. Using importance
sampling \citep{gordon1993novel}, the importance weights are computed
recursively as the ratio of the filtering distribution and the proposal
distribution $q$ (for the forward filter) and $\tilde{q}$ (for the
backward filter). Considering the auxiliary particle filter (APF)
of \citet{pitt1999filtering}, particles are proposed from the mixture
distribution,
\begin{equation}
q\left(\boldsymbol{\beta}_{j}|\mathbf{t}_{1:j}\right)=\sum_{k=1}^{K}\nu_{j}^{k}q\left(\boldsymbol{\beta}_{j}|\boldsymbol{\beta}_{j-1}^{k},\mathbf{t}_{j}\right),\label{eq: APF proposal}
\end{equation}
for the forward filter, and 
\begin{equation}
\tilde{q}\left(\tilde{\boldsymbol{\beta}}_{j}|\mathbf{t}_{j:J}\right)=\sum_{h=1}^{K}\tilde{\nu}_{j}^{h}\tilde{q}\left(\tilde{\boldsymbol{\beta}}_{j}|\tilde{\boldsymbol{\beta}}_{j+1}^{h},\mathbf{t}_{j}\right),\label{eq:APF backward proposal}
\end{equation}
for the backward filter. Where $\left\{ \nu_{j}^{h}\right\} _{h=1}^{K}$
and $\left\{ \tilde{\nu}_{j}^{k}\right\} _{h=1}^{K}$ are some normalized
mixture weights. To sample from these proposal distributions a mixture
component $a_{k}$ (referred to as ancestor) of $q$ is selected with
probability proportional to $\nu_{j}$ and then a particle $\boldsymbol{\beta}_{j}^{k}$
is proposed from $q(\boldsymbol{\beta}_{j}|\boldsymbol{\beta}_{j-1}^{a_{k}},\mathbf{t}_{j})$;
a similar procedure is used for $\tilde{q}$. Therefore, setting $\nu_{j}^{k}\wasypropto L_{j}(\mathbf{t}_{j}\text{|\ensuremath{\boldsymbol{\beta}}}_{j-1}^{k})w_{j-1}^{k}$
(and similarly $\tilde{\nu}_{j}^{h}\wasypropto L_{j}(\mathbf{t}_{j}\text{|\ensuremath{\beta}}_{j+1}^{h})\tilde{w}_{j+1}^{h}$
for the backward filter) implies that one proposes only from ancestors
that have high importance weights and high predictive density. 

Given the ancestors $a_{k}$ and $\tilde{a}_{h}$ from the forward
and backward filters respectively, the corresponding importance weights
become,
\begin{align}
w_{j}^{k} & \wasypropto\frac{p\left(\boldsymbol{\beta}_{j}^{k}|\,\mathbf{t}_{1:j}\right)}{q\left(\boldsymbol{\beta}_{j}^{k}|\,\mathbf{t}_{1:j}\right)}=\frac{L_{j}\left(\mathbf{t}_{j}|\,\boldsymbol{\beta}_{j}^{k}\right)p\left(\boldsymbol{\beta}_{j}^{k}|\,\boldsymbol{\beta}_{j-1}^{a_{k}}\right)}{L_{j}\left(\mathbf{t}_{j}\text{|\,\ensuremath{\boldsymbol{\beta}}}_{j-1}^{k}\right)q\left(\boldsymbol{\beta}_{j}^{k}|\,\boldsymbol{\beta}_{j-1}^{a_{k}},\mathbf{t}_{j}\right)},\nonumber \\
\tilde{w}_{j}^{h} & \wasypropto\frac{\tilde{p}\left(\tilde{\boldsymbol{\beta}}_{j}^{h}|\,\mathbf{t}_{j:J}\right)}{\tilde{q}\left(\tilde{\boldsymbol{\beta}}_{j}^{h}|\,\mathbf{t}_{j:J}\right)}=\frac{L_{j}\left(\mathbf{t}_{j}|\,\tilde{\boldsymbol{\beta}}_{j}^{h}\right)p\left(\tilde{\boldsymbol{\beta}}_{j+1}^{\tilde{a}_{h}}|\,\tilde{\boldsymbol{\beta}}_{j}^{h}\right)\gamma_{j}\left(\tilde{\boldsymbol{\beta}}_{j}^{h}\right)}{L_{j}\left(\mathbf{t}_{j}|\,\tilde{\boldsymbol{\beta}}_{j+1}^{\tilde{a}_{h}}\right)\widetilde{q}\left(\tilde{\boldsymbol{\beta}}_{j}^{h}|\,\tilde{\boldsymbol{\beta}}_{j+1}^{\tilde{a}_{h}},\mathbf{t}_{j}\right)\gamma_{j+1}\left(\tilde{\boldsymbol{\beta}}_{j+1}^{\tilde{a}_{h}}\right)}.\label{eq: IW forward and backward PF}
\end{align}

Finally, it remains to find a way of combining the two filters at
each time point $j$. Instead of evaluating the double mixture (\ref{eq:  particle smoothing distribution})
which is computationally costly, \citet{fearnhead2010sequential}
suggest running another particle filter that combines the samples
from the forward and backward particle filters recursively. This approach
relies on finding another proposal distribution $\bar{q}(\boldsymbol{\beta}_{j}|\boldsymbol{\beta}_{j-1}^{k},\mathbf{t}_{j},\tilde{\boldsymbol{\beta}}_{j+1}^{h})$
combining draws from the forward particle filter at the time point
$j-1$ and the backward particle filter at $j+1$. New smoothing particles
$\bar{\{\boldsymbol{\beta}_{j}}^{s}\}_{s=1}^{S}$are proposed from
$\bar{q}$, and the corresponding smoothing importance weights 
\begin{align}
\bar{w}_{j}^{s} & \propto\frac{p\left(\bar{\boldsymbol{\beta}_{j}}^{s}|\boldsymbol{\beta}_{j-1}^{k}\right)L_{j}\left(\mathbf{t}_{j}|\bar{\boldsymbol{\beta}_{j}}^{s}\right)p\left(\tilde{\boldsymbol{\beta}}_{j+1}^{h}|\,\bar{\boldsymbol{\beta}_{j}}^{s}\right)w_{j-1}^{k}\tilde{w}_{j+1}^{h}}{\bar{q}\left(\bar{\boldsymbol{\beta}}_{j}|\boldsymbol{\beta}_{j-1}^{k},\mathbf{t}_{j},\tilde{\boldsymbol{\beta}}_{j+1}^{h}\right)\nu_{j}^{k}\,\tilde{\nu}_{j}^{h}\,\gamma_{j+1}\left(\tilde{\boldsymbol{\beta}}_{j+1}^{h}\right)},\,s=1,\ldots S\nonumber \\
 & \propto\frac{p\left(\bar{\boldsymbol{\beta}_{j}}^{s}|\boldsymbol{\beta}_{j-1}^{k}\right)L_{j}\left(\mathbf{t}_{j}|\bar{\boldsymbol{\beta}_{j}}^{s}\right)p\left(\tilde{\boldsymbol{\beta}}_{j+1}^{h}|\,\bar{\boldsymbol{\beta}_{j}}^{s}\right)}{L_{j}\left(\mathbf{t}_{j}\text{|\,\ensuremath{\boldsymbol{\beta}}}_{j-1}^{k}\right)\bar{q}\left(\bar{\boldsymbol{\beta}}_{j}|\boldsymbol{\beta}_{j-1}^{k},\mathbf{t}_{j},\tilde{\boldsymbol{\beta}}_{j+1}^{h}\right)L_{j}\left(\mathbf{t}_{j}|\,\tilde{\boldsymbol{\beta}}_{j+1}^{h}\right)\gamma_{j+1}\left(\tilde{\boldsymbol{\beta}}_{j+1}^{h}\right)},\label{eq: smoothing IW}
\end{align}
approximate empirically the posterior distribution (\ref{eq:  particle smoothing distribution}).
The two-filter smoother algorithm of \citet{fearnhead2010sequential}
can be summarized in the following steps:

\begin{algorithm} \caption{The two-filter smoother} 
1. Forward particle filter \\
initialization: Sample $\beta_{0}^{k}\sim p\left(\beta_{0}\right)$ and set $w_{0}^{k}\propto \frac{1}{K}$\\

\For{$j = 1$ \KwTo $J$} { \vspace{1mm}
\For{$k = 1$ \KwTo $K$} { \vspace{1mm}

sample $\boldsymbol{\beta}_{j}^{k}\sim q(\boldsymbol{\beta}_{j}|\mathbf{t}_{1:j})$ and 
compute $w_{j}^{k}$ from (\ref{eq: IW forward and backward PF})
}
}
2. Backward particle filter \\
initialization: Sample $\beta_{J}^{h}\sim \gamma_{J}\left(\beta_{J}\right)$ and set $w_{J}^{k}\propto \frac{1}{K}$\\

\For{$j = J-1$ \KwTo $1$} { \vspace{1mm}
\For{$h = 1$ \KwTo $K$} { \vspace{1mm}

sample $\tilde{\boldsymbol{\beta}}_{j}^{h}\sim\tilde{q}(\boldsymbol{\beta}_{j}|\mathbf{t}_{j:J})$ and
compute $\tilde{w}_{j}^{h}$ from (\ref{eq: IW forward and backward PF})
}
}
3. Smoothing step: Combining the two filters \\
\For{$j = 1$ \KwTo $J$} { \vspace{1mm}
\For{$s = 1$ \KwTo $S$} { \vspace{1mm}

sample $\bar{\boldsymbol{\beta}_{j}}^{s}\sim\bar{q}(\boldsymbol{\beta}_{j}|\boldsymbol{\beta}_{j-1}^{k},\mathbf{t}_{j},\tilde{\boldsymbol{\beta}}_{j+1}^{h})$ and
compute $\bar{w}_{j}^{s}$ from (\ref{eq: smoothing IW})
}
}
\textbf{Output:}
$\{\bar{\boldsymbol{\beta}_{j}},\bar{w}_{j}\}_{j=1}^{J}$
\end{algorithm}

Here, there are two things to be noted. First, the sample size $S$
of the smoothing particles need not be equal to the sample size for
the forward (and backward) filtering particles $K$. One can gain
computation time by setting $K$ small and $S$ bigger than $K$.
More specifically by setting $S=RK$ ( in the following section $R=2$
in all runs of PSLiB). Second, the smoothing weights depend on the
artificial prior $\gamma_{j+1}$, which can be any distribution. In
order to ensure that the particles from backward filter are sampled
from the smoothing distribution, \citet{fearnhead2010sequential}
suggest setting $\gamma_{j+1}(\boldsymbol{\beta}_{j+1})=p(\boldsymbol{\beta}_{j+1}|\mathbf{t}_{1:j})$. 

\pagebreak{}

Given the particles sampled at the time point $j$ and their corresponding
importance weights, this prior can be represented by the mixture in
(\ref{eq:predictive prior approx}); however, doing this induces extra
computational costs. To avoid it, one can use the linear Bayes method
of \citet{west1985dynamic} to approximate the mixture by a single
Gaussian distribution (see Appendix \ref{sec:The-artificial-prior}).

\subsubsection{Proposal distribution based on linear Bayes method\label{subsec:Proposal-distribution}}

The aim of this section is to delineate the proposal distributions
$q$, $\tilde{q}$ and $\bar{q}$. Since the regression coefficients
are continuous random variables, it is convinient to construct $q$
as a mixture of Gaussian component distributions 
\begin{equation}
q(\boldsymbol{\beta}_{j}|\boldsymbol{\beta}_{j-1},\mathbf{t}_{j})\simeq N_{P+1}(\boldsymbol{\beta}_{j}|\mathbf{m}_{j},\mathbf{C}_{j}),\label{eq:Forward proposal approximation}
\end{equation}
where $\mathbf{m}_{j}$ and $\mathbf{C}_{j}$ are respectively the
mean and covariance matrix of the component distributions of the mixture
(\ref{eq: APF proposal}), and $P$ the dimension of the covariate
vector. The standard APF proposes particles from the random walk process
(\ref{eq:prior}), which means that $\mathbf{m}_{j}=\boldsymbol{\beta}_{j-1}$
and $\mathbf{C}_{j}=\mathbf{U}_{j}$. This choice is prone to high
degeneracy since particles are proposed from the prior. In addition
to that, the discount factor approach, which defines $\mathbf{U}_{j}=(\phi^{-1}-1)\boldsymbol{\Sigma}_{j-1}$,
makes this proposal impractical as it will be hard to control the
variance of the proposal. To obtain a better proposal distribution
it is necessary to include evidence from the data; one way to do this,
is to linearize the likelihood locally through a second order Tylor
series expansion of the log likelihood \citep{fearnhead2010sequential}
w.r.t the linear predictor $\eta_{j}$ around the mode value of $\eta_{j}$.
For the PEH model, the mode of the linear predictor for an individual
$i$ is $\hat{\eta}_{ij}=\log(\frac{d_{ij}}{t_{ij}})$ and the hessian
is $d_{ij}^{-1}$. Hence, this linearization is not guaranteed to
work since the mode and the hessian are not always finite -- both
$d_{ij}$ and $t_{ij}$ can be zero.

One alternative to get around this issue would be to first update
the parameter $\lambda_{j}$ (the index $i$ is omitted for notational
simplicity) and then exploit the fact that $\boldsymbol{\beta}_{j}$
enters the likelihood through the linear predictor, $\boldsymbol{\eta}_{j}=\ln\lambda_{j}=\mathbf{z}^{\prime}\mathbf{\boldsymbol{\beta}}_{j}$.
Since $\eta_{j}$ and $\boldsymbol{\beta}_{j}$ are connected deterministically,
one needs the posterior estimates $E[\eta_{j}|\mathbf{t}_{j}]$ and
$V[\eta_{j}|\mathbf{t}_{j}]$ to update the mean and variance of $q(\boldsymbol{\beta}_{j}|\boldsymbol{\beta}_{j-1},\mathbf{t}_{j})$
using the conditional expectations

\begin{align*}
\mathbf{m}_{j} & =E_{\eta_{j}}\left[E\left[\boldsymbol{\beta}_{j}|\eta_{j},\boldsymbol{\beta}_{j-1},\mathbf{t}_{1:j-1}\right]|\mathbf{t}_{j}\right],
\end{align*}
\begin{align}
\mathbf{C}_{j} & =E_{\eta_{j}}\left[V\left[\boldsymbol{\beta}_{j}|\eta_{j},\boldsymbol{\beta}_{j-1},\mathbf{t}_{1:j-1}\right]|\mathbf{t}_{j}\right]+\nonumber \\
 & \,\,\,V_{\eta_{j}}\left[E\left[\boldsymbol{\beta}_{j}|\eta_{j},\boldsymbol{\beta}_{j-1},\mathbf{t}_{1:j-1}\right]|\mathbf{t}_{j}\right].\label{eq:moments of PFP prop}
\end{align}
The inner expectation of (\ref{eq:moments of PFP prop}) are computed
from a joint (degenerate) prior of $\eta_{j}$ and $\boldsymbol{\beta}_{j}$,
\begin{equation}
\left(\begin{array}{c}
\eta_{j}\\
\boldsymbol{\beta}_{j}
\end{array}\right)|\boldsymbol{\beta}_{j-1},\mathbf{t}_{1:j-1}\sim N_{P+2}\left(\left(\begin{array}{c}
\boldsymbol{\mathbf{z}}^{\prime}\boldsymbol{\beta}_{j-1}\\
\boldsymbol{\beta}_{j-1}
\end{array}\right),\,\,\left[\begin{array}{cc}
\boldsymbol{\mathbf{z}}^{\prime}\hat{\boldsymbol{U}}_{j}\boldsymbol{\mathbf{z}} & \boldsymbol{\mathbf{z}}^{\prime}\mathbf{\hat{U}}_{j}\\
\hat{\boldsymbol{U}}_{j}\boldsymbol{\mathbf{z}} & \hat{\mathbf{U}}_{j}
\end{array}\right]\right).\label{eq:Degenerate prior}
\end{equation}
This method is known as linear Bayes and was proposed by \citet{west1985dynamic},
and is guaranteed to work as long as the posterior of $\lambda_{j}$
is twice differentiable with respect to $\eta_{j}$. For more details
see Appendix \ref{sec:Propal-distributions details}. 

Similarly, particles from the backward and the smoothing filters are
respectively proposed from 
\begin{align}
\tilde{q}\left(\boldsymbol{\beta}_{j}|\tilde{\boldsymbol{\beta}}_{j+1},\mathbf{t}_{j}\right) & \simeq N\left(\tilde{\boldsymbol{m}}_{j},\,\tilde{\boldsymbol{C}}_{j}\right),\nonumber \\
\bar{q}\left(\boldsymbol{\beta}_{j}|\boldsymbol{\beta}_{j-1},\mathbf{t}_{j},\tilde{\boldsymbol{\beta}}_{j+1}\right) & \simeq N\left(\bar{\boldsymbol{m}}_{j},\,\bar{\boldsymbol{C}}_{j}\right),\label{eq:smoothing proposal distribution}
\end{align}
where $\tilde{\boldsymbol{m}}_{j}=(1-\phi)\hat{\boldsymbol{\mu}}_{j}+\phi\tilde{\boldsymbol{\beta}}_{j+1}$,
$\tilde{\boldsymbol{C}}_{j}=(1-\phi)\hat{\boldsymbol{\Sigma}}_{j}$,
$\bar{\boldsymbol{m}}_{j}=(1-\phi)\boldsymbol{m}_{j}+\phi\tilde{\boldsymbol{\beta}}_{j+1}$
and $\boldsymbol{\bar{C}}_{j}=(1-\phi)\boldsymbol{C}_{j}$. $\hat{\boldsymbol{\mu}}_{j}$
and $\hat{\boldsymbol{\Sigma}}_{j}$ are defined in (\ref{eq:particle approx of the posterior estimate})
and $\boldsymbol{m}_{j}$ and $\boldsymbol{C}_{j}$ are the moments
of the proposal for the forward filter defined in (\ref{eq:moments of PFP prop});
further details are provided in Appendix \ref{sec:Details-on-the backward and smoothing proposals}.

\subsection{Model comparison and  prediction\label{subsec:Model-comparison-and prediction}}

The inference methodology developed here relies on three key model
choices: i) the discount factor, ii) the interval partition and iii)
the covariates used in the model. Inference for these elements is
based on the Watanabe-Akaike information creterion (WAIC); see \citet{gelman2014understanding}.
The WAIC uses the posterior sample and an out-of-sample test set to
estimate the log posterior predictive density with an adjustment for
the effective number of parameters. Given an out-of-sample test set
of size $n_{\textrm{test}}$
\begin{align}
WAIC & =\sum_{i=1}^{n_{\textrm{test}}}\log E_{\textrm{post}}\left[L\left(t_{i}^{*}|\boldsymbol{\beta}_{1:J}\right)\right]-V_{\textrm{post}}\left[\log L\left(t_{i}^{*}|\boldsymbol{\beta}_{1:J}\right)\right]\label{eq:WAIC}\\
 & =\sum_{i=1}^{n_{\textrm{test}}}\left\{ \log\left(E_{\textrm{post}}\left[L\left(t_{i}^{*}|\boldsymbol{\beta}_{1:J}\right)\right]\right)-E_{\textrm{post}}\left[\left(\log L\left(t_{i}^{*}|\boldsymbol{\beta}_{1:J}\right)-E_{\textrm{post}}\left[\log L\left(t_{i}^{*}|\boldsymbol{\beta}_{1:J}\right)\right]\right)^{2}\right]\right\} .\nonumber 
\end{align}

Given a sample of particles $\{\boldsymbol{\beta}_{1:J}^{k}\}_{k=1}^{K}$
from the smoothing posterior distribution and their corresponding
smoothing importance weights, $\{\bar{w}_{J}^{k}\}_{k=1}^{K}$ (note
that $\bar{w}_{J}$ are the weights for the complete paths $\boldsymbol{\beta}_{1:J}$),
the expectations in (\ref{eq:WAIC}) can be approximated as 
\begin{equation}
E_{\textrm{post}}\left[g\left(t_{i}|\beta_{1:J}\right)\right]=\frac{\sum_{k=1}^{K}g\left(t_{i}|\beta_{1:J}^{k}\right)\bar{w}_{J}^{k}}{\sum_{k=1}^{K}\bar{w}_{J}^{k}},\label{eq:IS approximation of expected values}
\end{equation}
where $g$ is any transformation of the likelihood function. Further,
the prediction of the probability that an individual $i$ survives
up to the time $t$ is approximated by the particle smoother as
\begin{align}
\hat{S}\left(t|\mathbf{x}_{i}^{*}\right) & =\frac{\sum_{k=1}^{K}\bar{w}_{J}^{k}\exp\left(-\left\{ \sum_{j=1}^{h-1}\left(\tau_{j}-\tau_{j-1}\right)\exp\left(\mathbf{z}_{i}^{*}{}^{\prime}\boldsymbol{\beta}_{j}^{k}\right)\right\} -\left(t-\tau_{h-1}\right)\exp\left(\mathbf{z}_{i}^{*}{}^{\prime}\boldsymbol{\beta}_{h}^{k}\right)\right)}{\sum_{k=1}^{K}\bar{w}_{J}^{k}},\label{eq: predicted survival}
\end{align}
if $\tau_{h-1}\leq t_{i}<\tau_{h},\,h\leq J$.

\section{Applications\label{sec:Applications}}

\subsection{Simulations\label{subsec:Simulations}}

A simulation study is conducted in order to assess the performance
of PSLiB in various senarios and compare it with the state-of-the-art
MCMC aproach in \citet{wagner2011bayesian}. The censoring indicators
$d_{i}$ ($i=1,\ldots,n$) are simulated from a Bernoulli distribution
with probability $1-p_{c}$ ($p_{c}$ being the proportion of censored
observations), and the exposure times, $t_{i}$, are simulated from
the following PEH models (using inverse sampling):
\[
\lambda_{ij}=\exp\left(\beta_{j,0}+\mathbf{x}_{i}^{\prime}\boldsymbol{\beta}_{j}\right),\,\,j=1,\ldots,J
\]
\[
\boldsymbol{\beta}_{j}=\boldsymbol{\beta}_{j-1}+\boldsymbol{\epsilon}_{j},\,\,\boldsymbol{\epsilon}_{j}\sim N_{P}\left(0,\,0.25I_{P}\right)
\]
\[
\beta_{j,0}=-11+\log\left(j\right),\,\,\boldsymbol{\beta}_{0}=0,\,x_{i}\sim N\left(0,I_{P}\right),
\]
where $\boldsymbol{\beta}_{j}=(\beta_{j,1},\ldots,\beta_{j,p})^{\prime}$
and $I_{p}$ is a $P\times P$ diagonal matrix.

The above data generating process (DGP) allows the flexibility of
varying the number of parameters $P$, the sample size $n$, the proportion
of the censored observations $p_{c}$ and the number of intervals
$J$ partitioning the study period. In all simulations, the intervals
are equidistant and have length equals to $20$ units of time.

The performance of the inference methodology is measured by the expected
discrimination measure (EDM) between the density for the DGP, $f(t|\mathbf{x})$,
and the predictive density $\hat{f}(t|\mathbf{x})$ of a given fitted
model, 
\[
EDM\left(f,\hat{f}\right)=\int\left[\int_{0}^{t}\frac{f\left(u|\mathbf{x}\right)}{F\left(t|\mathbf{x}\right)}\log\left(\frac{\nicefrac{f(u|\mathbf{x})}{F(t|\mathbf{x})}}{\nicefrac{\hat{f}(u|\mathbf{x})}{\hat{F}(u|\mathbf{x})}}\right)du\right]p\left(\mathbf{x}\right)d\mathbf{x},
\]
where $F(t|\mathbf{x})$ and $\hat{F}(t|\mathbf{x})$ are the cumulative
distribution functions (computed at time $t$) for the DGP and the
fitted model respectively, and $p(\mathbf{x})$ is the marginal distribution
of the covariates. For PEH models $F(t|\mathbf{x})=1-S(t|\mathbf{x})$
where $S(t|\mathbf{x})$ is the survival function. The expression
in the inner integral was proposed by \citet{di2004measure}, and
is referred to as the measure of discrimination between two past-life
distributions. To compute EDM, the inner intregral is evaluated numerically
using the trapezoidal rule and the outer integral is approximated
by taking the average over an out-of-sample test set of size $500$
simulated from $p(\mathbf{x})$. A value of EDM close to zero indicates
that the fitted model reconstructs the DGP very well.

\subsubsection{Performance assessment of PSLiB}

\begin{figure}[H]
\begin{raggedright}
\includegraphics[scale=0.45]{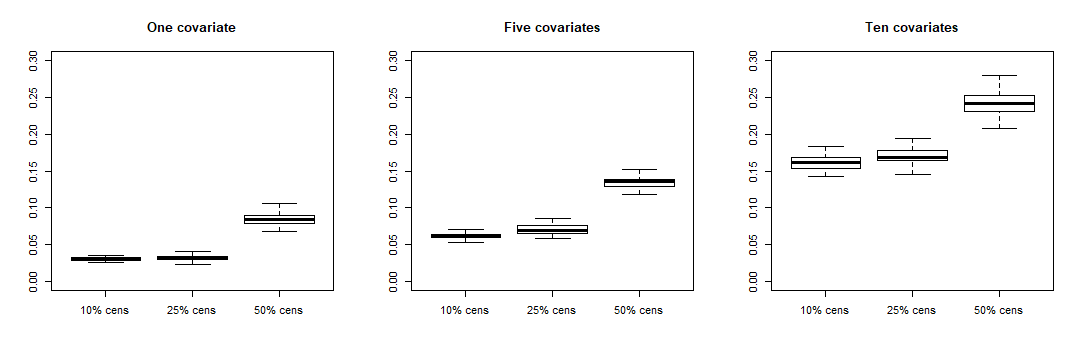}
\par\end{raggedright}
\caption{{\small{}Evaluating the performance of PSLiB with respect to the dimension
of the covariate vector and the proportion of censored observations.
Results are based on $50$ simulated datasets of size $n=2500$ and
$J=26$.}\label{Effect of censoring and size of the covariate}}
\end{figure}

To assess the effect of censoring and the covariate dimension on the
performance of PSLiB, datasets are simulated from different DGPs:
the number of covariates $P=1,5,10$, and the proportion of censored
observations $p_{c}=10\%,25\%,50\%$ (for each $P$). That is, in
total there are nine DGPs, and for each DGP, $50$ datasets of size
$n=2500$ and survival time length $J=26$ are simulated. Figure \ref{Effect of censoring and size of the covariate}
diplays the boxplots of the EDM between the DGP models and their corresponding
fitted models. Models are fitted using PSLiB with $K=2000$ particles.
Experiments show that increasing $K$ does not improve the EDM values.
Furthermore, a discount factor $0.3\leq\phi\leq0.45$ yields relatively
low EDM values with the lowest obtained in most cases when $\phi=0.45$
(results not shown). Therefore, in the following analyses $\phi$
is set to the latter value.

The EDM increases with both the number of variables and the proportion
of censored observations. For models with one covariate, the average
EDM is $0.030$, $0.032$ and $0.0850$ when the proportion of censored
observations is $10\%$, $25\%$ and $50\%$ respectively. For models
with five covariates, they increase to $0.061$, $0.071$ and $0.134$
respectively, and for models with ten covariates they increase even
more to $0.181$, $0.172$, and $0.253$ respectively.

Table (\ref{tab:Average-CPU-time}) presents the computation time
of the PSLiB algorithm for different sample sizes, number of covariates,
and number of intervals $J$. The effect of the number of particles
is not investigated since \citet{fearnhead2010sequential} has shown
that the particle smoother used here has a computational cost that
increases linearly with the number of particles.

\begin{table}[H]
\caption{{\small{}CPU time (in minutes) for training models on simulated data
with various sample sizes, number of covariates and length of the
survival time. Results are averages over $50$ simulated datasets
and $p_{c}=25\%$.}{\footnotesize{} }\label{tab:Average-CPU-time}}
\medskip{}

\centering{}%
\begin{tabular}{c|ccc|ccc|ccc|}
\hline 
\multirow{2}{*}{$\mathbf{n}$} & \multicolumn{3}{c|}{One covariate} & \multicolumn{3}{c|}{Five covariates} & \multicolumn{3}{c|}{Ten covariates}\tabularnewline
\cline{2-10} 
 & $\mathbf{J=6}$ & $\mathbf{J=16}$ & $\mathbf{J=26}$ & $\mathbf{J=6}$ & $\mathbf{J=16}$ & $\mathbf{J=26}$ & $\mathbf{J=6}$ & $\mathbf{J=16}$ & $\mathbf{J=26}$\tabularnewline
\hline 
$\mathbf{1000}$ & $0.20$ & $0.55$ & $0.92$ & $0.46$ & $1.36$ & $1.98$ & $0.52$ & $1.48$ & $2.08$\tabularnewline
$\mathbf{5000}$ & $0.32$ & $0.85$ & $1.29$ & $1.17$ & $2.98$ & $4.07$ & $1.24$ & $3.08$ & $4.10$\tabularnewline
$\mathbf{10000}$ & $0.47$ & $1.26$ & $1.69$ & $2.25$ & $5.45$ & $6.72$ & $2.36$ & $5.54$ & $7.01$\tabularnewline
\hline 
\end{tabular}
\end{table}
CPU time increases from $P=1$ to $P=5$, but then remains roughly
constant from $P=5$ to $P=10$. Therefore the number of covariates
in the model does not influence significantly the running time of
the algorithm. The effect of increasing $J$ is also modest: When
$J=26$, $p=10$ and $n=10,000$ it takes $7$ CPU minutes, but when
$n$ is increased to $50,000$ the computation time rises to $37$
CPU minutes and if $n$ is increased further to $100,000$ it takes$100$
CPU minutes to run the algorithm. Thus, for fixed covariate's dimension,
the computation time of PSLiB increases approximately linearly with
the sample size.

\subsubsection{Comparing PSLiB with AMS\label{subsec:Comparing-the-PSLiB}}

This section compares the PSLiB and the MCMC algorithm in \citet{wagner2011bayesian}
based on the number of effective sample size (ESS) of \citet{carpenter1999improved},
\[
ESS=\frac{\sigma_{j}^{2}}{E\left(\hat{\mu}_{j}-\mu_{j}\right)^{2}},
\]
where $\hat{\mu}_{j}$ is the posterior estimate of the regression
coefficients computed as in (\ref{eq:particle approx of the posterior estimate}),
and $\mu_{j}$ and $\sigma_{j}^{2}$ are respectively the true mean
and variance of the regression coefficient, which following \citet{carpenter1999improved},
are approximated by taking averages of $\hat{\mu}_{j}$ and $\hat{\sigma}_{j}^{2}$
over $M$ independent runs of the algorithm.

To compare both the ESS and computation time for PSLiB and AMS, ESS
per computation time (in CPU seconds) is used. The ESS is computed
based on $M=100$ independent replications of the PSLiB and AMS runs
on a dataset generated from a DGP with $P=1$, $J=26$ and $n=1000$.
The same number of iterations ($5000$ MCMC iteration/particles) is
set for both algorithms. The AMS model runs with the prior settings
proposed by \citet{wagner2011bayesian}.

\begin{figure}[H]
\begin{centering}
\includegraphics[scale=0.4]{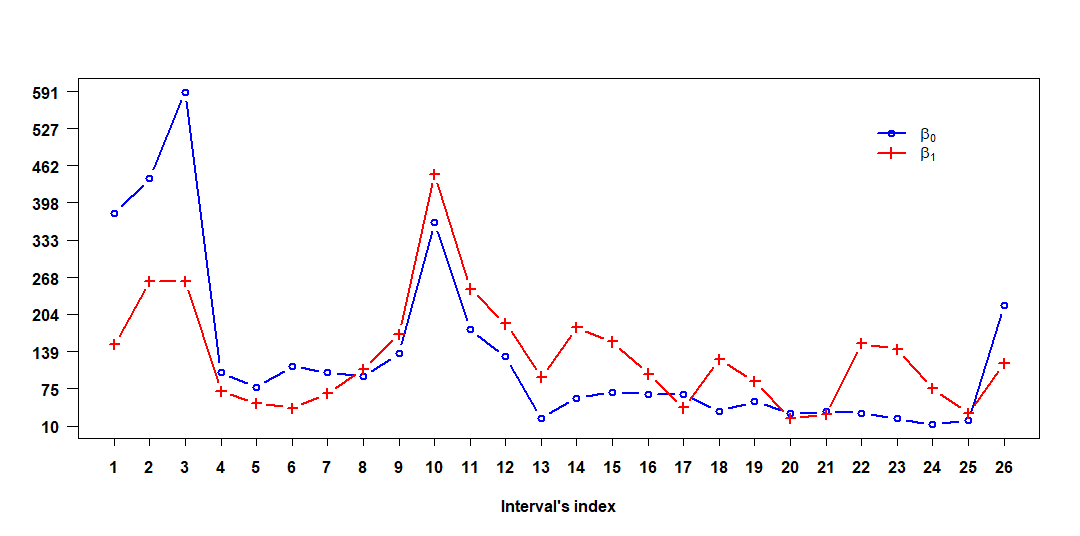}
\par\end{centering}
\caption{{\small{}Comparing the PSLiB and AMS algorithms using the ratio of
ESS/sec for PSLiB and AMS. The results are computated based on $100$
independent replications of each algorithm.\label{fig:Comparing PSLiB and the AMS}}}
\end{figure}
Figure \ref{fig:Comparing PSLiB and the AMS} displays the ratio of
the ESS/sec for PSLiB and AMS for all regression coefficients sampled
in all $J$ intervals. It is clear form Figure \ref{fig:Comparing PSLiB and the AMS}
that PSLiB is more efficient as it generates, on average, an effective
sample size that is more than two orders of magnitudes larger than
AMS (the average ratio is $133$ for $\beta_{0}$, and $131$ for
$\beta_{1}$). Comparing their computation time, PSLiB takes two minutes
(CPU time) while AMS takes roughly $105$ minutes. The main reason
for this difference is that PSLiB samples simultaneously all particles
at each time point $j=1,\ldots,J$ and also it does not require the
linearization of the likelihood in AMS.

\subsection{Patients with Acute Myocardial Infraction (AMI).}

In this section the PSLiB algorithm is applied to investigate the
effect of different risk factors on the survival time of patients
who were diagnosed with AMI. The dataset is a subset of the original
data analyzed by \citet{jensen1997does}, made available under the
name ``TRACE'' in the R package ``timereg'' \citep{scheike2009timereg}.
It contains survival times for $1878$ patients with AMI and six risk
factors: age in years (age), heart pumping measured in ultrasound
measurements (\textit{wmi}), ventricular fibrillation indicator --\textit{
vf} (\textit{$1$}: present, \textit{$0$}: absent), clinical heart
pump failure indicator --\textit{ chf }($1$: present, \textit{$0$}:
absent), indicator of diabetes ($1$: present, $0$: absent), and
sex ($1$: female, $0$: male). By the end of the follow-up time,
$970$ of the $1878$ patients ($52\%$) died from myocardial infraction
while the rest $908$ ($48\%$) were still alive or died from other
causes and, hence, were considered as censored. The aim of the study
is to estimate the effect of various risks on the survial time of
patients. 

The initial distribution, $p(\beta_{1})$, is set to the multivariate
normal distribution with mean zero, and a diagonal covariance matrix
where the variance of each covariate is set to $100$. The number
of particles is set to $10000$, and the continuous covariates (\textit{mwi}
and \textit{age}) are centered around their mean values. Since $970$
events are observed, it would be time consuming to partition the survivial
time based on the event times. Time is instead partitioned into $J$
intervals, such that each interval contains $E$ events; a lower $E$
value of course increases the number of intervals and the computation
time, but allows a more flexible form for the covariate effect. Note
that although all intervals contain the same number of events, they
need not be of the same length. 

The WAIC is used to compare models with different discount factors
$\phi$ and partitions of the survival time. The comparison is based
on a randomly selected test set of size $470$ (roughly $20\%$ of
the entire dataset) and the results are presented in Table \ref{tab:WAIC for different models}.

\begin{table}[H]
\caption{{\small{}Comparing models fitted with different discount factors (in
the rows) and different partitions of the study period (in columns)
based on the WAIC.}{\footnotesize{} }\label{tab:WAIC for different models}}
\medskip{}

\centering{}%
\begin{tabular}{c|cccc}
$\phi$ & \textbf{$\mathbf{E=20}$} & \textbf{$\mathbf{E=30}$} & \textbf{$\mathbf{E=40}$} & \textbf{$\mathbf{E=50}$}\tabularnewline
\hline 
\textbf{$\mathbf{0.9}$} & $2266.9$ & $2427.7$ & $2912.5$ & $3497.1$\tabularnewline
\textbf{$\mathbf{0.8}$} & $3775.7$ & $3063.9$ & $3256.1$ & $2549.7$\tabularnewline
\textbf{$\mathbf{0.7}$} & $3234.1$ & $2856.3$ & $2344.2$ & $1683.8$\tabularnewline
\textbf{$\mathbf{0.6}$} & $2139.4$ & $1659.5$ & $1420.8$ & $1357.4$\tabularnewline
\textbf{$\mathbf{0.5}$} & $1192.7$ & $1184.2$ & $1190.0$ & $1186.8$\tabularnewline
\textbf{$\mathbf{0.4}$} & $1278.3$ & $1232.5$ & $1214.9$ & $1202.5$\tabularnewline
\textbf{$\mathbf{0.3}$} & $1580.0$ & $1294.2$ & $1256.0$ & $1242.6$\tabularnewline
\hline 
\end{tabular}
\end{table}
The lowest WAIC indicating the best model is observed when $\phi=0.5$
and $E=30$. Therefore, for further analysis, the discount factor
and the number of events per interval are set respectively to the
selected values. A partition of $30$ events per interval results
into $32$ intervals; the initial intervals tend to be shorter than
the latter intervals due to the decreasing risk set over time.

\begin{figure}[H]
\begin{raggedright}
\includegraphics[scale=0.35]{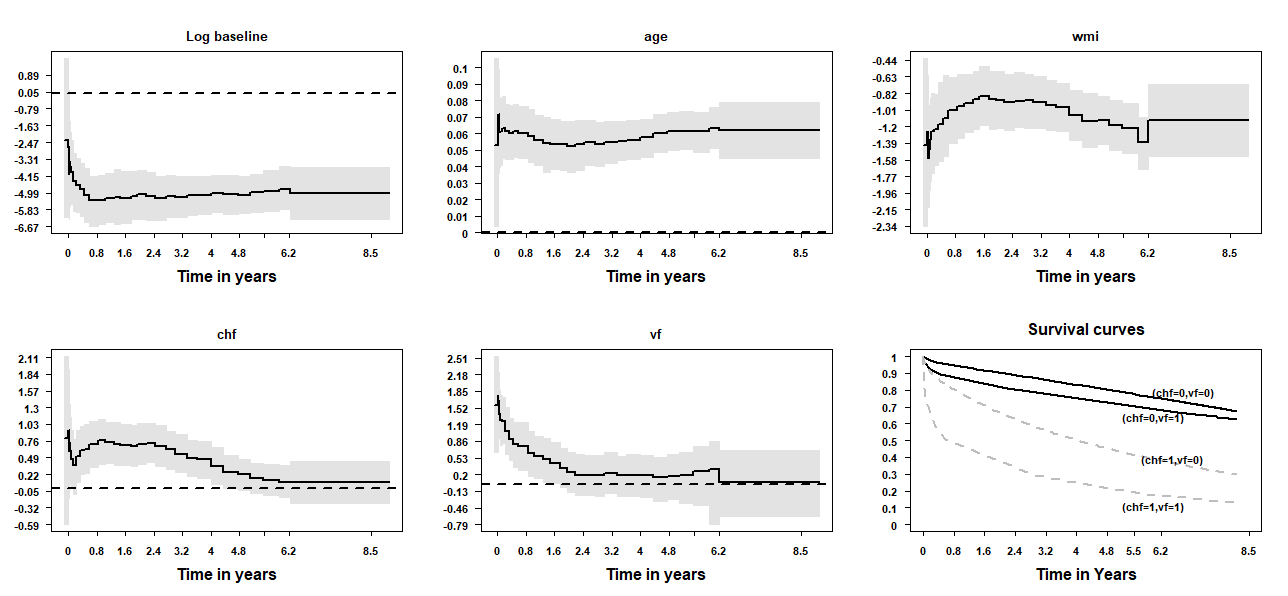}
\par\end{raggedright}
\caption{{\small{}Posterior mean trajectories of the log baseline hazards (intercept),
the effect of the risk factors with their corresponding $95\%$ credible
intervals (grey band), computed pointwise for each interval, and the
fitted survival curves. The horizontal dashed line is a reference
line at zero.}\label{fig:AMI-data:-Posterior}}
\end{figure}

The WAIC is also used to select the risk factors to be included in
further analysis of the AMI data. Setting the discount factor and
the number of event per interval to the above-mentioned selected values,
I fit models with all possible combination of the risk factors. The
best model (WAIC= $1181$) contains only the risk factors: \textit{age},
\textit{wmi}, \textit{chf} and \textit{vf}. The posterior estimates
for the parameters in the selected model as well as the predicted
survival curves are presented in Figure \ref{fig:AMI-data:-Posterior}.
The survival curve are computed by setting the continuous variables
(\textit{age} and \textit{wmi}) to their average values.

One can see that the risk factors \textit{wmi}, \textit{vf} and the
\textit{chf }have dynamic trends. The effect of the \textit{wmi} increases
in the first two and a half years then aftwerwards it starts declining
slowly, but later on after six years it stabilizes around $-1.01$
(corresponding to a relative hazard of $0.36$). However, it is always
negative indicating that the hazards of dying from AMI decreases with
increasing \textit{wmi}. The effect of \textit{chf }drops considerably
in the first month of diagnostics but later on it increases during
next few months of the first year. After that, it starts declining
slowly but always remains positive and converges to zero later on.
The effect of \textit{vf }is always positive as well but declines
monotonically and converges towards zero after five years. On the
other hand the effect of \textit{age} is nearly static since its mean
trajectory is more or less horizontal throughout the study period.
The risk of dying from AMI increases roughly at a rate of $6\%$ times
for each additional year on an individual's age. Furthermore, the
plot of the survival curves suggest that the group of patients with
$chf=0$ and $vf=0$ have the highest survival curve, followed by
the group with $chf=1$ and $vf=0$. The lowest survival curve corresponds
to the group with $chf=1$ and $vf=1$. Thus, the risk of dying from
AMI is much higher for patients with diagnosed with the heart pump
failure (\textit{chf}) compared to patients diagnosed with ventricular
fibrillation (\textit{vf}). The risk becomes even higher for patients
who have both \textit{chf} and \textit{vf}.

\section{Concluding remarks\label{sec:Concluding-remarks}}

An efficient algorithm for posterior inference in piecewise exponential
hazard models is proposed. The inference is based on a particle smoothing
algorithm which applies three particle filters to sample from the
posterior. This method requires designing efficient proposal distributions,
which are developed by approximating the posterior distribution of
the model parameters by a Gaussian distribution through a second order
Taylor series expansion and applying the linear Bayes method of \citet{west1985dynamic}.

The proposed inference methodology is shown to be fast and efficient
for relatively large and high dimensional data and it generates an
effective sample size that is more than two orders of magnitude higher
than the MCMC sampler of \citet{wagner2011bayesian}. Furthermore,
it has been applied to make inference on the effect of risk factors
of acute myocardial infraction and it turns out that the most important
risk factors (based on WAIC) are: \textit{age}, \textit{wmi}, \textit{chf}
and \textit{vf} , where the effect of all except the \textit{age}
vary with time. 

Possible further extensions of the present work would be to generalize
PSLiB to accomodates spatial covariates and/or time-varying covariates,
or to allow higher order splines for the regression coefficients. 

\bigskip{}

\textbf{\small{}Acknowledgements: }{\small{}I would like to thank
Mattias Villani, Gebrenegus Ghilagaber and Kevin Bruke for their constructive
comments, and Helga Wagner for sharing the code for the auxiliary
mixture sampler approach.}{\small\par}

\pagebreak{}

\bibliographystyle{chicago}
\addcontentsline{toc}{section}{\refname}\bibliography{Bibliography}

\pagebreak{}

\appendix

\section{The artificial prior distribution $\gamma_{j}$\label{sec:The-artificial-prior}}

The mean $\boldsymbol{\mu}_{j-1}$ and the covariance matrix $\boldsymbol{\Sigma}_{j-1}$
of the filtering distribution at time $j-1$ can be approximated from
realizations of the forward particle filter by:
\begin{equation}
\hat{\boldsymbol{\mu}}_{j-1}=\frac{\sum_{k=1}^{K}\boldsymbol{\beta}_{j-1}^{k}w_{j-1}^{k}}{\sum_{k=1}^{K}w_{j-1}^{k}},\,\,\,\hat{\boldsymbol{\Sigma}}_{j-1}=\frac{\sum_{k=1}^{K}\left(\boldsymbol{\beta}_{j-1}^{k}-\hat{\boldsymbol{\mu}}_{j-1}\right)\left(\boldsymbol{\beta}_{j-1}^{k}-\hat{\boldsymbol{\mu}}_{j-1}\right)^{\prime}w_{j-1}^{k}}{\sum_{k=1}^{K}w_{j-1}^{k}}\label{eq:particle approx of the posterior estimate}
\end{equation}

Therefore, the mean and the variance of the mixture (\ref{eq:predictive prior approx})
are approximated respectively by $\hat{\boldsymbol{\mu}}_{j-1}$ and
$\mathbf{\hat{R}}_{j}=\hat{\boldsymbol{\Sigma}}_{j-1}+\mathbf{\hat{U}}_{j}$,
where $\mathbf{\hat{U}}_{j}=(\phi^{-1}-1)\hat{\boldsymbol{\Sigma}}_{j-1}$
is the approximation of the variance of the random walk prior process
(\ref{eq:prior}) . It follows that the artificial prior, 
\begin{equation}
\gamma_{j}(\boldsymbol{\beta}_{j})\simeq N_{P+1}\left(\hat{\boldsymbol{\mu}}_{j-1},\frac{\hat{\boldsymbol{\Sigma}}_{j-1}}{\phi}\right).\label{eq:Artificial prior}
\end{equation}

\section{Details on the proposal distributions}

\subsection{The forward filter \label{sec:Propal-distributions details}}

Given the structure of the likelihood (\ref{eq:interval likelihood}),
the model parameter $\lambda_{ij}$ has a conjugate $\textrm{Gamma}(\alpha_{ij},\psi_{ij})$
prior distribution, which implies that the marginal posterior of $\lambda_{ij}$
is $\textrm{Gamma}(\alpha_{ij}+d_{ij},\psi_{ij}+t_{ij})$. Taking
into account the Jacobian of the transformation $\eta_{ij}=\ln\lambda_{ij}$,
it can be shown that

\begin{equation}
p\left(\eta_{ij}|\mathbf{t}_{1:j-1},\mathbf{t}_{1:i,j}\right)\propto\exp\left\{ \eta_{ij}\left(\alpha_{ij}+d_{ij}\right)-\left(\psi_{ij}+t_{ij}\right)\exp\left\{ \eta_{ij}\right\} \right\} ,\label{eq:marginal posterior of log hazard}
\end{equation}
where $\mathbf{t}_{1:i,j}$ is the set of exposure times for the first
$i$ individuals observed in the interval $I_{j}$. In order to apply
the conditional expectations (\ref{eq:moments of PFP prop}), a Laplace
approximation of the posterior (\ref{eq:marginal posterior of log hazard})

\[
N\left(\left[\dfrac{\partial\ln p\left(\eta_{ij}|\mathbf{t}_{1:j-1},\mathbf{t}_{1:i,j}\right)}{\partial\eta_{ij}}\right]_{\eta_{ij}=\hat{\eta_{ij}}},\left[-\dfrac{\partial^{2}\ln p\left(\eta_{ij}|\mathbf{t}_{1:j-1},\mathbf{t}_{1:i,j}\right)}{\partial\eta_{ij}^{2}}\right]_{\eta_{ij}=\hat{\eta_{ij}}}^{-1}\right)
\]
is required. Here $\hat{\eta_{ij}}$ is the mode value of the linear
predictor. The first and second derivatives in (\ref{eq:marginal posterior of log hazard})
are given by :

\[
\dfrac{\partial\ln p\left(\eta_{ij}|\mathbf{t}_{1:j-1},\mathbf{t}_{1:i,j}\right)}{\partial\eta_{ij}}=\alpha_{ij}+d_{ij}-\left(\psi_{ij}+t_{ij}\right)\exp\left\{ \eta_{ij}\right\} 
\]

\[
\dfrac{\partial^{2}\ln p\left(\eta_{ij}|\mathbf{t}_{1:j-1},\mathbf{t}_{1:i,j}\right)}{\partial\eta_{ij}^{2}}=-\left(\psi_{ij}+t_{ij}\right)\exp\left\{ \eta_{ij}\right\} 
\]

From the first derivative, one can show that the mode lies at $\hat{\eta_{ij}}=\ln(\frac{\alpha_{ij}+d_{ij}}{\psi_{ij}+t_{ij}})$,
which lead to the final expressions
\[
E[\eta_{j}|\mathbf{t}_{j}]=\ln\left(\frac{\alpha_{ij}+d_{ij}}{\psi_{ij}+t_{ij}}\right),\thinspace\thinspace V[\eta_{j}|\mathbf{t}_{j}]=\frac{1}{\alpha_{ij}+d_{ij}}
\]

The hyper-parameters $\alpha_{ij}$ and $\psi_{ij}$ are selected
in order to match the true moments of the prior with the moments from
the deterministic relationship $\eta_{ij}=\mathbf{z}_{i}^{\prime}\boldsymbol{\beta}_{j-1}$.
This is accomplished by setting $\ln\alpha_{ij}-\ln\psi_{ij}=\mathbf{z}_{i}^{\prime}\boldsymbol{\beta}_{j-1}$
and $\alpha_{ij}^{-1}=\mathbf{z}_{i}^{\prime}\mathbf{U}_{j}\mathbf{z}_{i}$;
hence $\psi_{ij}=\alpha_{ij}\exp\{-\mathbf{z}_{i}^{\prime}\boldsymbol{\beta}_{j-1}\}$.
The moments of the proposal $q$ described in Section \ref{subsec:Proposal-distribution}
are obtain from the recursive expressions (along $i=1,\ldots,n_{j}$),

\begin{align*}
\mathbf{m}_{ij} & =\boldsymbol{\beta}_{j-1}+\frac{\mathbf{A}_{ij}}{Q_{ij}}\left(\ln\left(\frac{\alpha_{ij}+d_{ij}}{\psi_{ij}+t_{ij}}\right)-a_{ij}\right)\\
 & =\boldsymbol{\beta}_{j-1}+\frac{\mathbf{A}_{ij}}{Q_{ij}}\ln\frac{1+Q_{ij}d_{ij}}{1+t_{ij}Q_{ij}\exp\left\{ a_{ij}\right\} }
\end{align*}

\begin{align*}
\mathbf{C}_{ij} & =\mathbf{U}_{j}-\frac{\mathbf{A}_{ij}\mathbf{A}_{ij}^{\prime}}{Q_{ij}}\left(1-\frac{1}{Q_{ij}\left(\alpha_{ij}+d_{ij}\right)}\right)\\
 & =\mathbf{U}_{j}-\mathbf{A}_{ij}\mathbf{A}_{ij}^{\prime}\left(\frac{d_{ij}}{1+d_{ij}Q_{ij}}\right)
\end{align*}

where $a_{ij}=\mathbf{z}_{i}^{\prime}\boldsymbol{\beta}_{j-1}$ ,
$\mathbf{A}_{ij}=\mathbf{U}_{j}\mathbf{z}_{i}$, and $Q_{ij}=\mathbf{z}_{i}^{\prime}\mathbf{U}_{j}\mathbf{z}_{i}$.
Starting with $i=1$, $\mathbf{m}_{1j}$ and $\mathbf{C}_{1j}$ are
computed with $\boldsymbol{\beta}_{j-1}$ drawn from the sample of
particles in the previous interval $I_{j-1}$ and then for $i+1$,
$\boldsymbol{\beta}_{j-1}=\mathbf{m}_{i,j}$ and $\mathbf{U}_{j}=\mathbf{C}_{ij}$.
Thus, $\mathbf{m}_{j}=\mathbf{m}_{n_{j},j}$ and $\mathbf{C}_{j}=\mathbf{C}_{n_{j},j}$.

\subsection{The backward and smoothing filters\label{sec:Details-on-the backward and smoothing proposals}}

Given the weigthed particle sample from the forward particle filter,
one can approximate $p(\boldsymbol{\beta}_{j}|\mathbf{z},\mathbf{t}_{1:j})$
by a Gaussian distribution with mean $\hat{\boldsymbol{\mu}}_{j}$
and covariance matrix $\hat{\boldsymbol{\Sigma}}_{j}$ defined in
(\ref{eq:particle approx of the posterior estimate}), the proposal
$\tilde{q}$ is derived from the following joint distribution 
\[
\left.\left(\begin{array}{c}
\boldsymbol{\beta}_{j}\\
\\
\boldsymbol{\beta}_{j+1}
\end{array}\right)\right|\mathbf{t}_{j},\mathbf{d}_{j}\sim N\left(\left(\begin{array}{c}
\hat{\boldsymbol{\mu}}_{j}\\
\\
\hat{\boldsymbol{\mu}}_{j}
\end{array}\right),\,\left(\begin{array}{ccc}
\hat{\boldsymbol{\Sigma}}_{j} &  & \hat{\boldsymbol{\Sigma}}_{j}\\
\\
\hat{\boldsymbol{\Sigma}}_{j} &  & \hat{R}_{j+1}
\end{array}\right)\right),
\]
where $\hat{R}_{j}$ is the variance of the prior $p(\boldsymbol{\beta}_{j+1}|\mathbf{t}_{j})$.
The discount factor approach of \citet{west1985dynamic} assumes that
$\hat{R}_{j+1}=\frac{\hat{\boldsymbol{\Sigma}}_{j}}{\phi}$ , $0<\phi<1$,
which results in the conditional moments
\[
E\left[\boldsymbol{\beta}_{j}|\boldsymbol{\beta}_{j+1},\mathbf{t}_{j}\right]=\hat{\boldsymbol{\mu}}_{j}+\phi\left(\boldsymbol{\beta}_{j+1}-\hat{\boldsymbol{\mu}}_{j}\right)=\phi\boldsymbol{\beta}_{j+1}+\left(1-\phi\right)\hat{\boldsymbol{\mu}}_{j}
\]

\[
V\left[\boldsymbol{\beta}_{j}|\boldsymbol{\beta}_{j+1},\mathbf{t}_{j}\right]=\hat{\boldsymbol{\Sigma}}_{j}-\phi\hat{\boldsymbol{\Sigma}}_{j}=\left(1-\phi\right)\hat{\boldsymbol{\Sigma}}_{j}
\]

Since (as shown in the derivation of $q$ ) $\mathbf{m}_{j}$ and
$\mathbf{C}_{j}$ correspond respectively to the mean and variance
of $p(\boldsymbol{\beta}_{j}|\boldsymbol{\beta}_{j-1},\mathbf{t}_{j})$,
therefore the moments of $\bar{q}$ are obtained by substituting $\mathbf{m}_{j}$
for $\hat{\boldsymbol{\mu}}_{j}$ and $\mathbf{C}_{j}$ for $\hat{\boldsymbol{\Sigma}}_{j}$
in the above expressions. Where, $\hat{\boldsymbol{\mu}}_{j}$ and
$\hat{\boldsymbol{\Sigma}}_{j}$ are computed according to (\ref{eq:Artificial prior}).
\end{document}